\documentclass[letterpaper,12pt]{article}
\usepackage{graphicx} 
\usepackage{multirow}
\usepackage{multicol}
\usepackage{latexsym}
\usepackage{amsmath}
\usepackage{amsfonts}
\usepackage{amssymb}
\usepackage{psfrag}
\usepackage{graphicx}
\usepackage{algorithm}
\usepackage[noend]{algpseudocode}
\usepackage[table]{xcolor}
\usepackage{colortbl}
\usepackage{tabularx}

\usepackage{multirow}
\begin{document}

\begin{center}
	
 \textbf{ {\fontsize{20}{60}\selectfont Data Integration through outcome adaptive LASSO and a collaborative propensity score approach}}\\

 Asma Bahamyirou $^1$*, Mireille E. Schnitzer $^1$
 \\
 
 $1$: Université de Montréal, Faculté de Pharmacie.
\end{center}

\begin{abstract}
Administrative data, or non-probability sample data, are increasingly being used to obtain official statistics due to their many benefits over survey methods. In particular, they are less costly, provide  a larger sample size, and are not reliant on the response rate. However, it is difficult to obtain an unbiased estimate of the population mean from such data due to the absence of design weights. Several estimation approaches have been proposed recently using an auxiliary probability sample which provides representative covariate information of the target population. However, when this covariate information is high-dimensional, variable selection is not a straight-forward task even for a subject matter expert. In the context of efficient and doubly robust estimation approaches for estimating a population mean, we develop two data adaptive methods for variable selection using the outcome adaptive LASSO and a collaborative propensity score, respectively.  Simulation studies are performed in order to verify the performance of the proposed methods versus competing methods.  Finally, we presented an anayisis of the impact of Covid-19 on Canadians.\\
Les sources de données administratives ou les échantillons non probabilistes sont de plus en plus considérés en pratique pour obtenir des statistiques officielles vu le gain qu’on en tire (moindre coût, grande taille d’échantillon, etc.) et le déclin des taux de réponse. Toutefois, il est difficile d’obtenir des estimations sans biais provenant de ces bases de données à cause du poids d’échantillonnage manquant. Des méthodes d’estimations ont été proposées récemment qui utilisent l’information auxiliaire provenant d’un échantillon probabiliste représentative de la population cible. En présence de données de grande dimension, il est difficile d'identifier les variables auxiliaires qui sont associées au mécanisme de sélection. Dans ce travail, nous développons une procédure de sélection de variables en utilisant le LASSO adaptatif et un score de propension collaboratif. Des études de simulations ont été effectuées en vue de comparer différentes approches de sélection de variables. Pour terminer, nous avons présenté une application sur l'impact de la COVID-19 sur les Canadiens.

\end{abstract}


\section{INTRODUCTION}

Administrative data, or non-probability sample data, are being increasingly used in practice to obtain official statistics due to their many benefits over survey methods (lower cost, larger sample size, not reliant on response rate).  However, it is difficult to obtain  unbiased estimates of population parameters from such data due to the absence of design weights. For example, the sample mean of an outcome in a non-probability sample would not necessarily represent the population mean of the outcome. Several approaches have been proposed recently using an auxiliary probability sample which provides representative covariate information of the target population. For example, one can estimate the mean outcome in the probability sample by using an outcome regression based approach. Unfortunately, this approach relies on the correct specification of a parametric outcome model. Valliant \& Dever (2011) used inverse probability weighting to adjust a volunteer web survey to make it representative of a larger population. Elliott \& Valliant (2017) proposed an approach to model the indicator representing inclusion in the nonprobability sample by adapting Bayes' rule.  Rafei et al. (2020) extended the Bayes' rule approach using Bayesian Additive
Regression Trees (BART). Chen (2016) proposed to calibrate non-probability samples using probability samples with the least absolute shrinkage and selection operator (LASSO). In the same context, Beaumont \& Chu (2020) proposed a tree-based approach for estimating the propensity score, defined as the probability that a unit belongs to the non-probability sample. Wisniowski et al. (2020) developed a Bayesian approach for integrating probability and nonprobability samples for the same goal. \\
Doubly robust
semiparametric methods such as the augmented inverse propensity weighted (AIPW) estimator (Robins, Rotnitzky and Zhao, 1994) and targeted minimum loss-based estimation (TMLE; van der Laan \& Rubin, 2006; van der Laan \& Rose, 2011) have been proposed to reduce the potential bias in the outcome regression based approach. The term doubly robust comes from the fact that these
methods require both the estimation of the propensity score model and the outcome
expectation conditional on covariates, where only one of which
needs to be correctly modeled to allow for consistent estimation of the parameter of interest.  Chen, Li \& Wu (2019) developed doubly robust inference with non-probability survey samples by adapting the Newton-Raphson procedure in this setting. Reviews and discussions of related approaches can be found in Beaumont (2020) and Rao (2020).

Chen, Li \& Wu (2019) considered the situation where the auxiliary variables are given, i.e. where the set of variables to include in the propensity score model is known. However, in practice or in high-dimensional data, variable selection for the propensity score may be required and it is not a straight-forward task even for a subject matter expert. In order to have unbiased estimation of the population mean, controlling for the variables that influence the selection into the non-probability sample and are also causes of the outcome is important (VanderWeele \& Shpitser, 2011). Studies have shown that including instrumental variables -- those that affect the selection into the non-probability sample but not the outcome -- in the propensity score model leads to inflation of the variance of the estimator relative to estimators that exclude such variables (Schisterman et al., 2009;
Schneeweiss et al., 2009; van der Laan \& Gruber, 2010). However, including variables that are only related to the outcome in the propensity score model will increase the precision of the estimator without affecting bias (Brookhart et al. 2006; Shortreed \& Ertefaie, 2017).
Using the Chen, Li \& Wu (2019) estimator for doubly robust inference with a non-probability sample, Yang, Kim \& Song (2020) proposed a two step approach for variable selection for the propensity score using the smoothly clipped absolute deviation (SCAD; Fan \& Li, 2001). Briefly, they used SCAD to select variables for the outcome model and the propensity score model separately. Then, the union of the two sets is taken to obtain the final set of the selected variables. To the best of our knowledge, their paper is the first to investigate a variable selection method in this context. 
In causal inference, multiple variable selection methods have been proposed for the propensity score model. We consider two in particular. Shortreed \& Ertefaie (OALASSO; 2017) developed the outcome adaptive LASSO. This approach uses the adaptive LASSO (Zou; 2006) but with weights in the penalty term that are the inverse of the estimated covariate coefficients from a regression of the outcome on the treatment and the covariates. Benkeser, Cai \& van der Laan (2019) proposed a collaborative-TMLE (CTMLE) that is robust to extreme values of propensity scores in causal inference.  Rather than estimating the true propensity score, this method instead fits a model for the probability of receiving the treatment (or being in the non-probability sample in our context) conditional on the estimated conditional mean outcome. Because the treatment model is conditional on a single-dimensional covariate, this approach avoids the challenges related to variable and model selection in the propensity score model. In addition, it relies on only sub-parametric rates of convergence of the outcome model predictions.

In this paper, we firstly propose a variable selection approach in high dimensional covariate settings by extending the outcome adaptive LASSO (Shortreed \& Ertefaie, 2017). The gain in the present proposal relative to the existing SCAD estimator (Yang, Kim \& Song 2020) is that the OALASSO can accommodate both the outcome and the selection mechanism in a one-step procedure. 
Secondly, we adapt the Benkeser, Cai \& van der Laan (2019) collaborative propensity score in our setting. Finally, we perform simulation studies in order to verify the performance of our two proposed estimators and compare them with the existing SCAD estimator for the estimation of the population mean. \\
The remainder of the article is organized as follows. In Section 2, we define our setting and describe our proposed estimators. In Section 3, we present the results of the
simulation study. We present an analysis of the impact of Covid-19 on Canadians in Section 4. A discussion is provided in Section
5.

\section{Methods}

In this section, we present the two proposed estimators in our setting: 1) an extension of the OALASSO for the propensity score (Shortreed \& Ertefaie, 2017) and 2) the application of Benkeser, Cai \& van der Laan's (2020) alternative propensity score.
\subsection{The framework}
\begin{figure}[H]
    \begin{center}
       \includegraphics[width=0.8\textwidth]{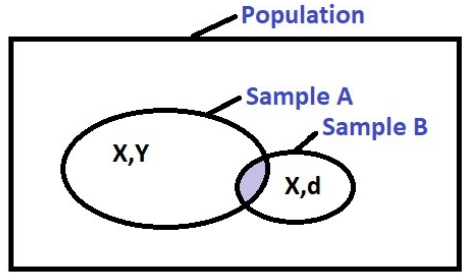}
    \caption{Population and observed samples.}
    \label{table:fig1}
    \end{center}
\end{figure}

Let $U=\{1,2,...,N\}$ be indices representing members of the target population. Define $\{\boldsymbol  X,Y\}$ as the auxiliary and response variables, respectively  where $\boldsymbol  X=(1, X^{(1)},X^{(2)},...,X^{(p)})$ is a vector of covariates (plus an intercept term) for an arbitrary individual. The finite target population data consists of $\{(\boldsymbol  X_i,Y_i), i\in U \}$. Let the parameter of interest be the finite population mean  $\mu=1/N\sum_{i\in U}Y_i$. Let $\mathcal{A}$ be indices for the non-probability sample  and let $\mathcal{B}$ be those of the probability sample. As illustrated in Figure \ref{table:fig1}, $\mathcal{A}$ and $\mathcal{B}$ are possibly overlapping subsets of $U$. Let $d_i=1/\pi_i$ be the design weight for unit $i$ with $\pi_i=P(i \in \mathcal{B})$ known. The data corresponding to $\mathcal{B}$ consist of observations $\{(\boldsymbol X_i,d_i): i \in \mathcal{B}\}$ with sample size $n_{\mathcal{B}}$. The data corresponding to the non-probability sample $\mathcal{A}$ consist of observations $\{(\boldsymbol X_i,Y_i): i \in \mathcal{A}\}$ with sample size $n_{\mathcal{A}}$. 

\begin{table}[H]
\centering
\begin{tabular}{||c c c c c c c||} 
 \hline
Sample& $X^{(1)}$ & ... & $X^{(p)}$ & Y & $\Delta$ & $d$\\ [0.5ex] 
 \hline\hline
\multirow{2}{*}{$\mathcal{A}$}&. &  ...  &  . &   . & 1 & \cellcolor{gray!40} \\ 
&. & ... & . & . & 1& \cellcolor{gray!40}\\
 
 \hline
 \multirow{2}{*}{$\mathcal{B}$} &. &  ...  &  . & \cellcolor{gray!40}   & 0 & .\\  [1ex] 
 &. & ... & . &\cellcolor{gray!40}  & 0 & .\\
 \hline
\end{tabular}
\caption{Observed data structure}
\label{table:tab1}
\end{table}

The observed data (Table \ref{table:tab1}) can be
represented as $O = \{\boldsymbol X, \Delta, I_{\mathcal{B}}, \Delta Y, I_{\mathcal{B}} d\}$, where  $\Delta$ is the indicator which equals 1 if
the unit belongs to the non-probability sample $\mathcal{A}$ and 0 otherwise, $I_{\mathcal{B}}$ is the indicator which equals 1 if
the unit belongs to the probability sample $\mathcal{B}$ and 0 otherwise, and $d$ is the design weight.  We use $O_i= \{\boldsymbol X_i, \Delta_i, I_{\mathcal{B},i}, \Delta_i Y_i, I_{\mathcal{B},i} d_i\}$
to represent the i-th subject's data realization. Let $p_i=P(\Delta_i=1|\boldsymbol X_i)$ be the propensity score (the probability of
the unit belonging to $\mathcal{A}$). In order to identify the target parameter, we
assume these conditions in the finite population: (1) Ignorability, such that the selection indicator $\Delta$ and the response variable $Y$ are independent given the set of covariates $\boldsymbol X$ (i.e. $\Delta \perp  Y | \boldsymbol X$) and (2) positivity such that $p_i>\epsilon>0$ for all $i$. Note that assumption (1)  implies that $E(Y|\boldsymbol X)=E(Y|\boldsymbol X,\Delta=1)$, which means that the conditional expectation of the outcome can be estimated using only the non-probability sample $\mathcal{A}$. Assumption (2) guarantees that all units have a non-zero probability of belonging in the non-probability sample.

\subsection{Estimation of the propensity score}\label{sec_pscores}
Let's assume for now that the propensity score follows a logistic regression model with $p_i=p(\boldsymbol X_i,\boldsymbol \beta)=exp(\boldsymbol X^T_i\boldsymbol\beta)/\{1+exp(\boldsymbol X^T_i\boldsymbol\beta)\}$. The true parameter value $\boldsymbol \beta_0$ is defined as the argument of the minimum (arg min) of the risk function $\sum_{i=1}^{N}[ \Delta_i \log\{p(\boldsymbol X_i,\boldsymbol\beta)\}+(1-\Delta_i) \log\{1-p(\boldsymbol X_i,\boldsymbol\beta)\}] $, with summation taken over the target population. One can rewrite this based on our observed data as 
\begin{equation}\label{eq1}
    \boldsymbol \beta_0      =  \arg \min_{\boldsymbol \beta} \sum_{\mathcal{A}} \boldsymbol X_i^T\boldsymbol\beta+ \sum_{i=1}^{N} \log(1+e^{\boldsymbol X_i^T\boldsymbol\beta}). 
\end{equation}
Equation (\ref{eq1}) cannot be solved directly since $\boldsymbol X$ has not been observed for all units
in the finite population. However, using the design weight of the probability sample $\mathcal{B}$, $\boldsymbol \beta_0$ can be estimated by minimising the pseudo risk function as 
\begin{equation}\label{eq22}
 \arg \min_{\boldsymbol \beta} \sum_{\mathcal{A}} \boldsymbol X_i^T\boldsymbol\beta- \sum_{\mathcal{B}}d_i\log(1+e^{\boldsymbol X_i^T\boldsymbol\beta}).
\end{equation}
Let $\boldsymbol X_{\mathcal{B}}$ be the matrix of auxiliary information (i.e. the design matrix) of the sample $\mathcal{B}$ and $\mathcal{L}(\boldsymbol\beta)$ the pseudo risk function defined above. Define $U(\boldsymbol\beta) = \frac{\partial \mathcal{L}(\boldsymbol\beta)}{\partial \beta} =\sum_{\mathcal{A}}  \boldsymbol X_i- \sum_{\mathcal{B}}p_id_i \boldsymbol X_i$, the gradient of the pseudo risk function.
Also define $H(\boldsymbol\beta) =  -\sum_{\mathcal{B}}d_ip_i(1-p_i) \boldsymbol X_i \boldsymbol X_i^T = \boldsymbol X^T_{\mathcal{B}}\boldsymbol S_{\mathcal{B}} \boldsymbol X_{\mathcal{B}}$, the Hessian of the pseudo risk function, where $S_i= -d_i p_i(1-p_i)$ and vector $\boldsymbol{S}_{\mathcal{B}}=(S_i; i\in \mathcal{B})$. The parameter $\boldsymbol \beta$ in equation (\ref{eq22}) can be obtained by solving the Newton-Raphson iterative procedure as proposed in Chen, Li \& Wu (2019) by setting $\boldsymbol \beta^{(t+1)} = \boldsymbol \beta^{(t)}-{H\{\boldsymbol \beta^{(t)}\}}^{-1}U\{\boldsymbol \beta^{(t)}\}$.
\subsubsection{Variable selection for propensity score}\label{sec_ALASSO}
In a high dimensional setting, suppose that an investigator would like to choose relevant auxiliary variables for the propensity score that could help to reduce the selection bias and standard error when estimating the finite population mean. In the causal inference context of estimating the average treatment effect, Shortreed \& Ertefaie (2017) proposed the OALASSO to select amongst the $X^{(j)}$s in the propensity score model. They penalized the aforementioned risk function by the adaptive LASSO penalty (Zou, 2006) where the coefficient-specific weights are the inverse of an estimated outcome regression coefficient representing an association between the outcome, $Y$, and the related covariate. 

In our setting, let the true coefficient values of a regression of $Y$ on $\boldsymbol X$ be denoted ${\alpha}_j$. The parameters $\boldsymbol\beta=(\beta_0,\beta_1, ... ,\beta_p)$, corresponding to the covariate coefficients in the propensity score, can be estimated by minimizing the pseudo risk function in (\ref{eq22}) penalized by the adaptive LASSO penalty:
\begin{equation}\label{eq2}
 \widehat{\boldsymbol \beta} = \arg \min_{\boldsymbol \beta} \sum_{\mathcal{A}} \boldsymbol X_i^T\boldsymbol\beta- \sum_{\mathcal{B}}d_i\log(1+e^{\boldsymbol X_i^T\boldsymbol\beta})  +\lambda\sum_{j=1}^{p}\check{\omega}_j |\beta_j|.
\end{equation}
where $\check{\omega}_j = 1 / |\check{\alpha}_j|^{\gamma}$ for some $\gamma > 0 $ and $\check{\alpha}_j$ is a $\sqrt{n}$-consistent estimator of $\alpha_j$.\\
Consider a situation where variable selection is not needed ($\lambda=0$). Chen, Li \& Wu (2019) proposed to estimate $\boldsymbol\beta$ by solving the Newton-Raphson iterative procedure. One can rewrite the gradient as $U(\boldsymbol\beta)  = \sum_{\mathcal{B}} [\Delta_i d_i-p_id_i] \boldsymbol X_i=\boldsymbol X^T_{\mathcal{B}}(\boldsymbol{\Sigma}_{\mathcal{B}} - \boldsymbol{Z}_{\mathcal{B}})$ with vectors $\boldsymbol{Z}_{\mathcal{B}}=(p_id_i; i\in \mathcal{B})$ and $\boldsymbol{\Sigma}_{\mathcal{B}}= (\Delta_i d_i; i\in \mathcal{B})$.

The Newton-Raphson update step can be written as:
\begin{equation}\label{eq4}
    \begin{array}{lll}
\boldsymbol \beta^{(t+1)}  & = & \boldsymbol\beta^{(t)} - (\boldsymbol X^T_{\mathcal B}\boldsymbol S_{\mathcal B}\boldsymbol X_{\mathcal B})^{-1}\boldsymbol X^T_{\mathcal B}(\boldsymbol{\Sigma}_{\mathcal{B}} - \boldsymbol{Z}_{\mathcal{B}})\\
       & = & (\boldsymbol X^T_{\mathcal B}\boldsymbol S_{\mathcal B} \boldsymbol X_{\mathcal B})^{-1}\boldsymbol X^T_{\mathcal B}\boldsymbol S_{\mathcal B} Y^*
\end{array} 
\end{equation}
where  $Y^* =\boldsymbol X_{\mathcal{B}}\boldsymbol\beta^{(t)}- \boldsymbol S_{\mathcal B}^{-1}(\boldsymbol{\Sigma}_{\mathcal{B}} - \boldsymbol{Z}_{\mathcal{B}})$. Equation (\ref{eq4})  is equivalent to the estimator of the weighted least squares problem with $Y^*$ as the new working response and $S_i= -d_i p_i(1-p_i)$ as the weight associated with unit $i$. Thus, in our context as well, we can select the important variables in the propensity score by solving a weighted least squares problem penalized with an adaptive LASSO penalty.
\subsubsection{Implementation of OALASSO}\label{sec_estps}
Now we describe how our proposal can be easily implemented
in a two-stage procedure. In the first stage, we construct the pseudo-outcome by using the Newton-Raphson estimate of $\boldsymbol\beta$ defined in equation (\ref{eq22}) and the probability sample $\mathcal{B}$. In the second stage, using sample $\mathcal{B}$, we solve a weighted penalized least squares problem with the pseudo-outcome as response variable. The selected variables correspond to the non-zero coefficients of the
adaptive LASSO regression.
The proposed algorithm for estimating the parameters $\boldsymbol\beta$ in equation (\ref{eq2})
with a given value of $\lambda$ is as follows:
\begin{algorithm}[H]
\caption{OALASSO for propensity score estimation}\label{alg:euclid}
\begin{algorithmic}[1]
\State Use the  Newton-Raphson algorithm for the unpenalized logistic regression in Chen, Li \& Wu (2019) to estimate $\tilde{\boldsymbol\beta}$ in (\ref{eq22}).
\State Obtain the estimated propensity score $\tilde{p}_i=p(\boldsymbol X_i,\tilde{\boldsymbol\beta})$ for each unit.
\State Construct an estimate of the new working response $Y^*$ by plugging in the estimated $\tilde{\boldsymbol\beta}$.
\State  Select the useful variables by following steps (a)-(d) below: 
\begin{itemize}
  \item[(a)] Define $S_i = -d_i \tilde{p}_i(1-\tilde{p}_i)$ for each unit in $\mathcal{B}$.
  \item[(b)] Run a parametric regression of $Y$ on $\boldsymbol X$  using sample $\mathcal{A}$. Obtain $\check{\alpha}_j$, the estimated coefficient of $X^{(j)}$, $j=1,...,p$.
  \item[(c)] Define the adaptive LASSO weights $\check{\omega}_j = 1/ |\check{\alpha_j}|^{\gamma}$, $j=1,...,p$ for $\gamma > 0$.
  \item[(d)]  Using sample $\mathcal{B}$, run a LASSO regression of $Y^*$  on $\boldsymbol X$ with $\check{\omega}_j$ as the penalty factor associated with $X^{(j)}$ with the given $\lambda$.
    $$\widehat{\boldsymbol\beta} = \arg \min_{\boldsymbol\beta} \sum_{\mathcal B} S_i(Y^*_i-\boldsymbol\beta^T\boldsymbol X_i)^2 +\lambda \sum_{j=1}^p  \check{\omega}_j |\beta_j|$$
    \item[(e)] The non-zero coefficient estimate from (d) are the selected variables.
\end{itemize}
\State The final estimate of the propensity score is $\widehat{p}_i=p(\boldsymbol X_i,\widehat{\boldsymbol\beta})=\frac{exp(\boldsymbol X^T_i\widehat{\boldsymbol\beta})}{1+exp(\boldsymbol X^T_i\widehat{\boldsymbol\beta})}$
\end{algorithmic}
\end{algorithm}
For the adaptive LASSO tuning parameters, we choose $\gamma=1$ (Nonnegative Garotte Problem; Yuan  \& Lin, 2007) and $\lambda$ is selected using V-fold cross-validation in the sample $\mathcal B$. The sampling design needs to be taken into account when creating the V-fold in the same way that we form random groups for variance estimation (Wolter, 2007). For cluster or stratified sampling for example, all elements in the cluster or stratum should be placed in the same fold.

\subsubsection{SCAD variable selection for propensity score}\label{sec_Yang}
Yang, Kim \& Song (2020) proposed a two step approach for variable selection using SCAD. In the first step, they used SCAD to select relevant variables for both the propensity score and the outcome model, respectively. Denote $\mathcal{C}_p$ (respectively $\mathcal{C}_m$) the selected set of relevant variables for the propensity score (respectively the outcome model). The final set of variables used for estimation is $\mathcal{C} = \mathcal{C}_p 	\cup \mathcal{C}_m$.

\subsection{Inverse weighted estimators}
Horvitz \& Thompson (1952) proposed the idea of weighting observed values by inverse probabilities of selection in
the context of sampling methods. The same idea is used to estimate the population mean in the missing outcome setting. Recall that $p(\boldsymbol X) = P(\Delta=1| \boldsymbol X)$ is the propensity score. 
In order to estimate the population mean, the units in the non-probability sample $\mathcal{A}$ are assigned the weights $w_i = 1/ \widehat{p}_i$ where $\widehat{p}_i = p(\boldsymbol X_i,\widehat{\boldsymbol\beta})$ is the estimated propensity score obtained using Algorithm 1. The inverse probability weighted (IPW) estimator for the population mean is given by
$$ \mu_n^{IPW}= \sum_{i\in \mathcal{A}} w_i Y_i / \widehat{N},$$

where $\widehat{N} = \sum_{i \in \mathcal{A}} w_i$. For the estimation of the variance, we use the proposed variance of Chen, Li \& Wu (2019) which is given by
$$\widehat{V}(\mu_n^{IPW})=\frac{1}{\widehat{N}_{\mathcal{A}}^2}\sum_{i\in \mathcal{A}} (1-\widehat{p}_i)\left(\frac{Y_i-\mu_n^{IPW}}{\widehat{p}_i}-\boldsymbol{\widehat{b}}_2^T\boldsymbol X_i\right)^2 + \widehat{\boldsymbol b}_2^T \widehat{\boldsymbol D} \widehat{\boldsymbol b}_2$$ 
with $\widehat{\boldsymbol D}=\widehat{N}_{\mathcal{B}}^{-2}  V_p(\sum_{i\in \mathcal{B}}d_i\widehat{p}_i\boldsymbol X_i)$, where $V_p(·)$ denotes the design-based variance of the total
under the probability sampling design for $\mathcal{B}$ and 
$$\widehat{\boldsymbol b}_2 = \left\{\sum_{i\in \mathcal{A}} \left(\frac{1}{\widehat{p}_i} -1\right)(Y_i-\mu_n^{IPW})\boldsymbol X_i^T \right\}\left\{ \sum_{i\in \mathcal{B}}d_i\widehat{p}_i(1-\widehat{p}_i) \boldsymbol X_i\boldsymbol X_i^T \right\}^{-1} $$
\subsection{Augmented Inverse Probability Weighting}
Doubly robust semi-parametric methods such as AIPW (Scharfstein, Rotnitzky \&
Robins, 1999) or Targeted Minimum Loss-based Estimation (TMLE, van der Laan \& Rubin, 2006; van der Laan \& Rose, 2011)
have been proposed to potentially reduce the error resulting from misspecified outcome regressions but also avoid total dependence on the propensity score model
specification. We denote $m(\boldsymbol X)=E(Y|\boldsymbol X)$ and let $\widehat{m}(\boldsymbol X)$ be an estimate of $m(\boldsymbol X)$. 
   Under the current setting, the AIPW estimator proposed in Chen, Li \& Wu (2019) for $\mu$ is
$$ \mu_n^{AIPW}=\frac{1}{\widehat{N}_{\mathcal{A}}} \sum_{i\in \mathcal{A}}\frac{ Y_i-\widehat{m}(\boldsymbol X_i)}{p(\boldsymbol X_i,\widehat{\boldsymbol\beta})} +\frac{1}{\widehat{N}_{\mathcal{B}}} \sum_{i\in \mathcal{B}}d_i\widehat{m}(\boldsymbol X_i)$$
where $\widehat{N}_{\mathcal{A}} = \sum_{i\in \mathcal{A}}1/p(\boldsymbol X_i,\widehat{\boldsymbol\beta})$, $\widehat{N}_{\mathcal{B}} = \sum_{i\in \mathcal{B}}d_i$ and $\widehat{\boldsymbol\beta}$ can be estimated using either the Newton-Raphson algorithm in Chen, Li \& Wu (2019) or our proposed OALASSO. One can also use the alternative propensity score proposed by Benkeser, Cai \& van der Laan (2020) and therefore replacing $\widehat{p}_i = p(\boldsymbol X_i,\widehat{\boldsymbol\beta})$ by the estimated probability of belonging to the nonprobability sample conditional on the estimated outcome regression $\widehat{P}\{\Delta=1|\widehat{m}(\boldsymbol X_i)\}$.\\
 For the estimation of the variance, we use the proposed variance of Chen, Li \& Wu (2019) which is given by
$$\widehat{V}(\mu_n^{AIPW})=\frac{1}{\widehat{N}_{\mathcal{A}}^2}\sum_{i\in \mathcal{A}} (1-\widehat{p}_i)\left\{\frac{Y_i-\widehat{m}(\boldsymbol X_i)-\widehat{H}_N}{\widehat{p}_i}-\widehat{\boldsymbol b}_3^T\boldsymbol X_i \right\}^2 + \widehat{W}$$
with  $\widehat{H}_N =\widehat{N}_{\mathcal{A}}^{-1}\sum_{i\in \mathcal{A}} \{Y_i-\widehat{m}(\boldsymbol X_i)\}/\widehat{p}_i$, $\widehat{t}_i = \widehat{p}_i \boldsymbol X_i^T \widehat{\boldsymbol b}_3 + \widehat{m}(\boldsymbol X_i)-\widehat{N}_{\mathcal{B}}^{-1} \sum_{i\in \mathcal{B}}d_i\widehat{m}(\boldsymbol X_i)$, $\widehat{W}=1/\widehat{N}_{\mathcal{B}}^2  V_p(\sum_{i\in \mathcal{B}}d_i\widehat{t}_i)$, where $V_p(·)$ denotes the design-based variance of the total
under the probability sampling design for $\mathcal{B}$ and 
$  \widehat{\boldsymbol b}_3 = \left[\sum_{i\in \mathcal{A}} \left(\frac{1}{\widehat{p}_i} -1\right)\{ Y_i-\widehat{m}(\boldsymbol X_i)-\widehat{H}_N \}\boldsymbol X_i^T \right]\left\{ \sum_{i\in \mathcal{B}}d_i\widehat{p}_i(1-\widehat{p}_i) \boldsymbol X_i\boldsymbol X_i^T \right\}^{-1}. $

\section{Simulation study}
\subsection{Data generation and parameter estimation}
We consider a similar simulation setting as Chen, Li \& Wu (2019). However, we
add 40 pure binary noise covariates (unrelated to the selection mechanism or outcome) to our set of covariates. We generate a finite population $\mathcal{F}_N=\{(\boldsymbol X_i,Y_i):i=1,...,N\}$ with $N=10,000$, where $Y$ is the outcome variable and $\boldsymbol X=\{X^{(1)},...,X^{(p)}\}, p=44$ represents the auxiliary variables.  Define $Z_1 \sim Bernoulli(0.5)$, $Z_2\sim Uniform(0,2)$, $Z_3 \sim Exponential(1)$ and $Z_4 \sim \chi^2(4)$. The observed outcome $Y$ is a Gaussian with a mean  $\theta = 2 + 0.6X^{(1)} + 0.6X^{(2)} + 0.6X^{(3)} + 0.6X^{(4)}$, where $X^{(1)}=Z^{(1)}$, $X^{(2)}= Z^{(2)}+0.3X^{(1)}$, $X^{(3)}=Z^{(3)} +0.2\{X^{(1)}+X^{(2)}\}$, $X^{(4)}=Z^{(4)} + 0.1\{X^{(1)}+X^{(2)}+X^{(3)}\}$, with  $X^{(5)},...,X^{(24)} \sim Bernoulli(0.45)$ and $X^{(25)},...,X^{(44)} \sim N(0,1)$.\\ From the finite population, we select a probability sample  $\mathcal{B}$ of size $n_{\mathcal{B}} \approx 500$ under a Poisson sampling with probability $\pi \propto \{0.25 + X^{(2)}+0.03Y\}$. We also consider three scenarios for selecting a non-probability sample $\mathcal{A}$  with the inclusion indicator $\Delta \sim Bernoulli(p)$:
\begin{itemize}
    \item Scenario 1 considers a situation in which the confounders $X^{(1)}$ and $X^{(2)}$ (common causes of inclusion and the outcome) have a weaker relationship with inclusion ($\Delta=1$) than with the outcome:
      $P(\Delta=1|\boldsymbol X)=\textrm{expit}\{-2+ 0.3 X^{(1)}+ 0.3 X^{(2)}-X^{(5)}-X^{(6)}\}$
    \item Scenario 2 considers a situation in which both confounders $X^{(1)}$ and $X^{(2)}$ have a weaker relationship with the outcome  than with inclusion:
    $P(\Delta=1|\boldsymbol X)=\textrm{expit}\{-2+  X^{(1)}+  X^{(2)}-X^{(5)}-X^{(6)}\}$
    \item Scenario 3 involves a stronger association between the instrumental variables $X^{(5)}$ and $X^{(6)}$ and inclusion:  $P(\Delta=1|\boldsymbol X)=\textrm{expit}\{-2+ X^{(1)}+X^{(2)}-1.8X^{(5)}-1.8X^{(6)}\}$
\end{itemize}
To evaluate the performance of our method in a nonlinear setting (Scenario 4), we simulate a fourth setting  following exactly Kang \& Schafer (2007). In this scenario, we generate independent $Z^{(i)}$ $ \sim N(0,1), i=1,..4$. The observed outcome is generated as $Y = 210 +27.4 Z^{(1)} + 13.7 Z^{(2)} + 13.7 Z^{(3)}+13.7 Z^{(4)} + \epsilon$, where $\epsilon \sim N(0,1)$ and the true propensity model is $P(\Delta=1 \mid\boldsymbol Z)=\textrm{expit}\{-Z^{(1)}+ 0.5 Z^{(2)} - 0.25 Z^{(3)} - 0.1 Z^{(4)}\}$. However, the analyst observes the variables $X^{(1)} = \textrm{exp}\{Z^{(1)}/2\}$, $X^{(2)}=Z^{(2)}/[1+\textrm{exp}\{Z^{(1)}\}] +10$, $X^{(3)} = \{Z^{(1)}Z^{(3)}/25 + 0.6\}^3$, and $X^{(4)}= \{Z^{(2)} + Z^{(4)}+20\}^2$ rather than the $Z^{(j)}$s.\\
The parameter of interest is the population mean $\mu_0=N^{-1}\sum_{i=1}^{N}Y_i$. Under each scenario, we use a correctly specified outcome regression model for the estimation of $m(\boldsymbol X)$. For
the estimation of the propensity score, we perform logistic regression with all $44$ auxiliary variables as main terms, LASSO, and OALASSO, respectively. For the Benkeser method, we also use logistic regression for the propensity score.
Because the 4th scenario involves model selection but not variable selection, we only compare logistic regression with the Benkeser method for the propensity score. We fit a misspecified model and the highly adaptive LASSO (Benseker \& van der Lann, 2016) for the outcome model. \\

The performance of each estimator is evaluated
through the percent bias ($\%B$), the mean
squared error (MSE) and the coverage rate (COV), computed as
\begin{align*}
\%B = \frac{1}{R}\sum_{r=1}^{R} \frac{\widehat{\mu}_r-\mu}{\mu} \times 100\\
MSE =\frac{1}{R}\sum_{r=1}^{R} (\widehat{\mu}_r-\mu)^2\\
COV = \frac{1}{R}\sum_{r=1}^{R} I(\mu \in \widehat{CI}_r)
\end{align*} respectively, where $\widehat{\mu}_r$ is the estimator computed from the $r$th simulated
sample, $\widehat{CI}_r=(\widehat{\mu}_r -1.96\sqrt{v_r}, \widehat{\mu}_r +1.96\sqrt{v_r}) $ is the confidence interval with $v_r$ the estimated variance using the method proposed by Chen, Li \& Wu (2019) for the $r$th
simulation sample, and $R = 1000$ is the total number of simulation runs.

\subsection{Results}
Tables \ref{table:scenario1}, \ref{table:scenario2} and \ref{table:scenario3} contain the results for the first three scenarios. In all three, the IPW estimators performed the worst overall in terms of \% bias. Similar to Chen, Li \& Wu (2019), the coverage rates of IPW were suboptimal in all scenarios and the standard error was substantially underestimated.   The AIPW estimator, implemented with logistic regression, LASSO and OALASSO for the propensity score, performed very well in all scenarios with unbiased estimates and coverage rates close to the nominal $95\%$. In comparison to IPW and AIPW with logistic regression, incorporating the LASSO or the OALASSO did not improve the bias but did lower the variance and allowed for better standard error estimation. The Benkeser method  slightly increased the bias of AIPW and had underestimated standard errors, leading to lower coverage. The Yang method  had the highest bias compared to the other implementations of AIPW and greatly overestimated standard error in all three scenarios.

\begin{figure}[H]
    \begin{center}
       \leavevmode
       \includegraphics[width=1.2\textwidth]{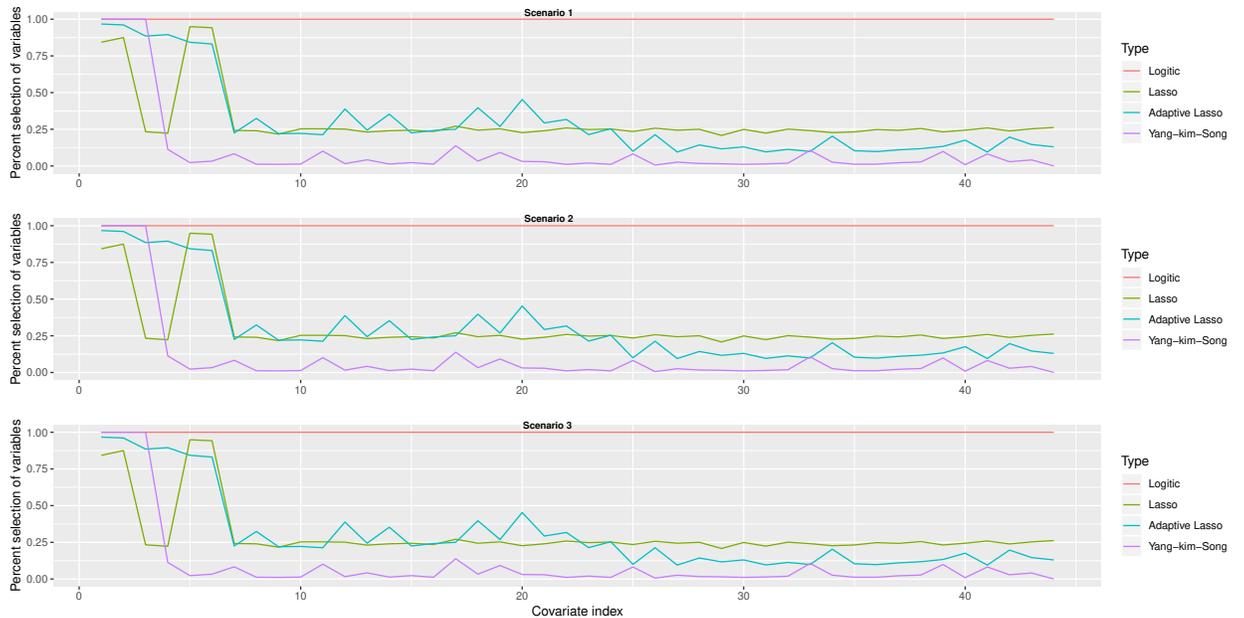}
    \caption{Percent selection of each variable into the propensity score model over $1000$ simulations  under scenarios 1-3.}
    \label{table:fig2}
    \end{center}
\end{figure}

For the first three scenarios, Figure \ref{table:fig2} displays the percent selection of each covariate (1,...,44), defined as the percentage of estimated coefficients that are non-zero throughout
the 1000 generated datasets. Overall, the LASSO tended to select the true predictors of inclusion: $X^{(1)}$, $X^{(2)}$, $X^{(5)}$ and $X^{(6)}$. For example, in scenario (2), confounders ($X^{(1)}$, $X^{(2)}$) were selected in around $94\%$ of simulations and instruments ($X^{(5)}$, $X^{(6)}$) around $90\%$. However, the percent selection of pure causes of the outcome ($X^{(3)}$, $X^{(6)}$) was around $23\%$. On the other hand, when OALASSO was used for the propensity score, the percent selection  of confounders ($X^{(1)}$, $X^{(2)}$) was around $98\%$ and instruments ($X^{(5)}$, $X^{(6)}$) was $64\%$. However, the percent selection of pure causes of the outcome ($X^{(3)}$, $X^{(4)}$) increased to $83\%$. When using Yang's proposed selection method, $X^{(1)}$, $X^{(2)}$ and $X^{(3)}$ were selected $100$ percent of the time.\\
Table \ref{table:KangShafer} contains the results of the Kang and Shafer (2007) setting. AIPW with HAL for the outcome model and either the collaborative propensity score (AIPW - Benkeser method) or propensity score with logistic regression with main terms (AIPW - Logistic (2) ) achieved lower \% bias and MSE compared to IPW. However, when the outcome model was misspecified, AIPW with logistic regression (AIPW - Logistic (1) ) performed as IPW.  In this scenario, the true outcome expectation and the propensity score functionals were nonlinear, making typical parametric models misspecified. Consistent estimation of the outcome expectation can be obtained by using flexible models. 
The collaborative propensity score was able the reduce the dimension of the space and collect the necessary information using the estimated conditional mean outcome for unbiased estimation of the population mean with a coverage rate that was close to nominal.

\begin{table}[H]
\small\sf\centering
\caption{Scenario 1: Estimates taken over
1000 generated datasets. $\%$B (percent bias), MSE (mean squared error), MC SE (monte carlo standard error), SE (mean standard error) and COV (percent coverage). IPW-Logistic: IPW with logistic regression for propensity score; IPW-LASSO: IPW with LASSO regression for propensity score; IPW-OALASSO: IPW with OALASSO regression for propensity score; AIPW-Logistic: AIPW with logistic regression for propensity score; AIPW-LASSO: AIPW with LASSO regression for propensity score; AIPW-OALASSO: AIPW with OALASSO regression for propensity score; AIPW-Benkeser: AIPW with the collaborative propensity score; AIPW-Yang: Yang's proposed AIPW.} 
\label{table:scenario1} 
\begin{tabular}{l c c c cc} 
\hline
\textbf{Estimator} & \textbf{$\%B$} & \textbf{MSE} & \textbf{MC SE} & \textbf{SE} & \textbf{$\%$COV}\\
\hline 

 IPW - Logistic &-0.51  &  0.03 &0.17 & 0.12 & 79\\
 IPW - LASSO &  -1.75 & 0.03&0.12 & 0.07 & 49\\
 IPW - OALASSO & -0.94 &  0.06 & 0.25 & 0.07 & 42\\
 AIPW - Logistic  & 0.01 & 0.01& 0.10 & 0.14  & 99  \\
 AIPW - LASSO  & -0.03 & 0.01& 0.10& 0.09 & 93\\
 AIPW - OALASSO  & -0.01 & 0.01& 0.10& 0.11 & 94  \\
  AIPW - Benkeser  & -0.00 &  0.01& 0.10&0.08  & 90 \\
 AIPW - Yang & -0.60 &  0.01 & 0.10& 0.30 &100\\

\hline 
\end{tabular}
\end{table}

\begin{table}[H]
\small\sf\centering
\caption{Scenario 2: Estimates taken over
1000 generated datasets. $\%$B (percent bias), MSE (mean squared error), MC SE (monte carlo standard error), SE (mean standard error) and COV (percent coverage). IPW-Logistic: IPW with logistic regression for propensity score; IPW-LASSO: IPW with LASSO regression for propensity score; IPW-OALASSO: IPW with OALASSO regression for propensity score; AIPW-Logistic: AIPW with logistic regression for propensity score; AIPW-LASSO: AIPW with LASSO regression for propensity score; AIPW-OALASSO: AIPW with OALASSO regression for propensity score; AIPW-Benkeser: AIPW with the collaborative propensity score; AIPW-Yang: Yang's proposed AIPW.} 
\label{table:scenario2} 
\begin{tabular}{l c c c cc} 
\hline
\textbf{Estimator} & \textbf{$\%B$} & \textbf{MSE} & \textbf{MC SE} & \textbf{SE} & \textbf{$\%$COV}\\
\hline 

 IPW - Logistic & 0.74  &  0.03 & 0.18 & 0.09 & 66 \\
 IPW - LASSO &  -2.49 & 0.03&0.11 &0.04 & 25 \\
 IPW - OALASSO & -1.32  &  0.04&  0.18 & 0.05& 37\\
 AIPW - Logistic  & 0.03 & 0.01 & 0.10 & 0.18  & 100\\
 AIPW - LASSO  & -0.08 &  0.01& 0.10  & 0.09 & 94 \\
 AIPW - OALASSO  & -0.04 &  0.01& 0.10 & 0.10  & 95\\
 AIPW - Benkeser  &  -0.13 &   0.01& 0.10  & 0.06 & 83\\
 AIPW - Yang & -1.59 &  0.02 &  0.09&  0.29 & 100 \\

\hline 
\end{tabular}
\end{table}

\begin{table}[H]
\small\sf\centering
\caption{Scenario 3: Estimates taken over
1000 generated datasets. $\%$B (percent bias), MSE (mean squared error), MC SE (monte carlo standard error), SE (mean standard error) and COV (percent coverage). IPW-Logistic: IPW with logistic regression for propensity score; IPW-LASSO: IPW with LASSO regression for propensity score; IPW-OALASSO: IPW with OALASSO regression for propensity score; AIPW-Logistic: AIPW with logistic regression for propensity score; AIPW-LASSO: AIPW with LASSO regression for propensity score; AIPW-OALASSO: AIPW with OALASSO regression for propensity score; AIPW-Benkeser: AIPW with the collaborative propensity score; AIPW-Yang: Yang's proposed AIPW.} 
\label{table:scenario3} 
\begin{tabular}{l c c c cc} 
\hline
\textbf{Estimator} & \textbf{$\%B$} & \textbf{MSE} & \textbf{MC SE} & \textbf{SE} & \textbf{$\%$COV}\\
\hline 

 IPW - Logistic & 1.11  &  0.13& 0.36 & 0.29 &  87\\
 IPW - LASSO &  -3.01 & 0.06 &0.15 & 0.09 &  44\\
 IPW - OALASSO &  -1.71  &  0.07&0.25  & 0.11 & 58\\
 AIPW - Logistic  & 0.03 & 0.02& 0.15 & 0.32 & 100\\
 AIPW - LASSO  & 0.03 &   0.01& 0.10 & 0.09 &  93\\
 AIPW - OALASSO  &  0.03  &  0.01& 0.10 & 0.10& 94\\
 AIPW - Benkeser  & 0.06  &   0.01& 0.10 & 0.07 &85\\
 AIPW - Yang & -1.59 &  0.02 & 0.10 & 0.30 &100\\
\hline 
\end{tabular}
\end{table}

\begin{table}[H]
\small\sf\centering
\caption{Scenario 4 (non-linear model setting): Estimates taken over
1000 generated datasets. $\%$B (percent bias), MSE (mean squared error), MC SE (monte carlo standard error), SE (mean standard error) and COV (percent coverage). IPW-Logistic: IPW with logistic regression for propensity score; AIPW-Logistic (1): AIPW with logistic regression for propensity score and a misspecified model for the outcome; AIPW-Logistic (2): AIPW with logistic regression for propensity score and HAL for the outcome model; AIPW-Benkeser: AIPW with the collaborative propensity score.} 
\label{table:KangShafer} 
\begin{tabular}{l c c c cc} 
\hline
\textbf{Estimator} & \textbf{$\%B$} & \textbf{MSE} & \textbf{MC SE} & \textbf{SE} & \textbf{$\%$COV}\\
\hline 
IPW - Logistic & 3.32  &  66.56 & 0.40 & 0.34& 0\\
AIPW - Logistic (1)  & 3.33 & 66.56 & 0.40 & 1.15 & 0 \\
 AIPW - Logistic (2)  & 0.12 & 2.60 & 1.59 & 1.00&86\\
 AIPW - Benkeser  & 0.12 &   2.61  &  1.59 & 1.28&93\\
\hline 
\end{tabular}
\end{table}

\section{Data Analysis}
In this section, we apply our proposed method to a survey which was conducted by Statistics Canada to measure the impacts of COVID-19 on Canadians. The main topic was to determine the level of trust Canadians have in others (elected officials, health authorities, other people, businesses and organizations) in the context of the COVID-19 pandemic. Data was collected from May 26 to June 8, 2020. The dataset was completely non-probabilistic
with a total of $35,916$ individuals responding and a wide range of basic demographic information collected from participants along with the main topic variables. The dataset is referred to as Trust in Others (TIO).\\
We consider Labor Force Survey (LFS) as a reference dataset, which
consists of $n_{\mathcal{B}}=89,102$ subjects with survey weights. This dataset does not have measurements of the study outcome variables of interest; however, it contains a rich set of auxiliary information common with the TIO. Summaries (unadjusted sample means for TIO and design-weighted means for LFS) of the common covariates are  listed in Tables \ref{table:char1} and \ref{table:char2} in the appendix. It can be seen that the distributions of the common covariates between the two samples are different. Therefore, using TIO only to obtain any estimate about the Canadian population may be subject to selection bias.\\
We apply the proposed methods and the sample mean to estimate the population mean of two response variables. Both of these variables were assessed as ordinal : $Y_1$, ``trust in decisions on reopening, Provincial/territorial government" -- 1: cannot be trusted at all, 2, 3: neutral, 4, 5: can be trusted a lot; and $Y_2$, ``when a COVID-19 vaccine becomes available, how likely is it that you will choose to get it?" -- 1: very likely, 2: somewhat likely, 3: somewhat unlikely, 4: very unlikely, 7: don't know. $Y_1$ was converted to a binary outcome which equals $1$ for a value less or equal to $3$ (neutral) and $0$ otherwise. The same type of conversion was applied for $Y_2$ to be $1$ for a value less or equal to 2 (somewhat likely) and $0$ otherwise. We used logistic regression, outcome adaptive group LASSO (Wang \& Leng, 2008;  Hastie et al. 2008; as we have categorical covariates), and the Benkeser method for the propensity score. We also fit group LASSO for the outcome regression when implementing AIPW. Each categorical covariate in Table \ref{table:char1},\ref{table:char2} were converted to binary dummy variables. Using $5$-fold cross-validation, the group LASSO variable selection procedure identified all available covariates in the propensity score model. Table \ref{table:application} below presents the point estimate, the standard error and the $95\%$ Wald-type confidence intervals. For estimating the standard error, we used the variance estimator for IPW and the asymptotic variance for AIPW proposed in Chen, Li \& Wu (2019). For both outcomes, we found significant differences in estimates between the naive sample mean and our proposed methods for both AIPW with OA group LASSO and the Benkeser method. For example, the adjusted estimates for $Y_1$ suggested that, on average, at most $40\%$ (using both outcome adaptive group LASSO or the Benkeser method) of the Canadian population have no trust at all or are neutral in regards to decisions on reopening taken by their provincial/territorial government compared to $43\%$ if we would have used the naive mean. The adjusted estimates for $Y_2$ suggested that  at most $80\%$ using the Benkeser method (or $82\%$ using outcome adaptive group LASSO)  of the Canadian population are very or somewhat likely to get the vaccine compared to $83\%$ if we would have used the naive mean. In the othe hand, there was no significant differences between OA group LASSO and group LASSO compared to the naive estimator. The package IntegrativeFPM (Yang, 2019) threw errors during application, which is why it is not included.

\begin{table}[H]
\small\sf\centering
\caption{Point estimate, standard error and $95\%$  Wald confidence interval. IPW-Logistic (Grp LASSO/OA Grp LASSO): IPW with logistic regression (Group LASSO/outcome adaptive Group LASSO) for propensity score; AIPW-Logistic (Grp LASSO/OA Grp LASSO): AIPW with logistic regression (Group LASSO/outcome adaptive Group LASSO for propensity score;  AIPW-Benkeser: AIPW with the collaborative propensity score.} 
\label{table:application} 
\begin{tabular}{l c c c} 
\hline
\multirow{6}{4em}{ $Y_1$ } & Sample mean  & 0.430 (0.002)  & 0.424 - 0.435 \\
&  IPW - Logistic  & 0.382 (0.024)  & 0.330 - 0.430 \\
& IPW - Grp LASSO  & 0.383 (0.024)  & 0.335 - 0.431 \\
& IPW - OA Grp LASSO  & 0.386 (0.024)  & 0.340 - 0.433 \\
& AIPW - Logistic    & 0.375 (0.022) & 0.328 - 0.415 \\
& AIPW - Grp LASSO   & 0.372 (0.014)  &  0.344 - 0.401 \\
& AIPW - OA Grp LASSO   & 0.373 (0.014)  &  0.348 - 0.403 \\
& AIPW - Benkeser  & 0.401 (0.002) &  0.396 - 0.406 \\
 \hline
\multirow{6}{4em}{ $Y_2$ }  &  Sample mean  & 0.830 (0.001) & 0.826 - 0.834 \\
&  IPW - Logistic  &  0.820 (0.013)  &  0.794 - 0.847 \\
& IPW - Grp LASSO  &  0.810 (0.013)  &  0.784 - 0.836 \\
& IPW - OA Grp LASSO  &  0.808 (0.013)  &  0.784 - 0.833 \\
&  AIPW - Logistic   & 0.810 (0.013) & 0.784 - 0.837 \\
&  AIPW - Grp LASSO   & 0.796 (0.012)  &  0.774 - 0.819 \\
&  AIPW - OA Grp LASSO   & 0.796 (0.011)  &  0.775 - 0.818 \\
& AIPW - Benkeser & 0.788 (0.003)  & 0.783 - 0.794 \\
\hline 
\end{tabular}
\end{table}

\section{Discussion}
In this paper, we proposed an approach to  variable selection for propensity score estimation through penalization when combining a non-probability sample with a reference probability sample. We also illustrated the application of the collaborative propensity score method of  Benkeser, Cai \& van der Laan (2020) with AIPW in this context. Through the simulations, we studied the performance of the different estimators and compared them with the method proposed by Yang. We showed that the LASSO and the OALASSO can reduce the standard error and mean squared error in a high dimensional setting. The collaborative propensity score produced good results but the related confidence intervals were suboptimal as the true propensity score is not estimated there.\\ Overall, in our simulations, we have seen that doubly robust estimators generally outperformed the IPW estimators. Doubly robust estimators incorporate the outcome expectation in such a way that can help to reduce the bias when the propensity score model is not correctly specified. Our observations point to the importance of using doubly robust methodologies in this context.\\
In our application, we found statistically significant differences in the results between
our proposed estimator and the corresponding naive estimator for both outcomes. This analysis used the variance estimator proposed by  Chen, Li \& Wu (2019) which relies on the correct specification of the propensity score model for IPW estimators. For future research, it would
be quite interesting to develop a variance estimator that is robust to propensity score misspecification and that can be applied to the Benkeser method. Other possible future directions include post-selection variance estimation in this setting.

\section*{ACKNOWLEDGEMENTS}
This work was supported by Statistics Canada and the Natural Sciences and Engineering Research Council of Canada
(Discovery Grant and Accelerator Supplement to MES), the Canadian Institutes of Health Research
(New Investigator Salary Award to MES) and the Facult\'e de pharmacie at Universit\'e de Montr\'eal
(funding for AB and MES). The authors thank
Jean-Francois Beaumont (Statistics Canada) for his very helpful comments on the manuscript.\\
{\it Conflict of Interest}: None declared.\\

\begin{appendix}

\begin{table}[H]
\caption{\textbf{Distributions of common covariates from the two samples. }}
\label{table:selec}
\begin{tabular}{l c c c c c c c}
\hline
 Methods & $X^{(1)}$ & $X^{(2)}$  & $X^{(3)}$ & $X^{(4)}$ & $X^{(5)}$ & $X^{(6)}$ & $X^{(7),..,(44)}$  \\
\hline
&  \multicolumn{7}{c}{\bf Scenario 1} \\
LASSO & 25 &34  & 13 & 11 &  67&  65 & 24\\
OALASSO  & 67  &  65 & 57  & 56 & 42 & 40 & 20 \\
Yang's method & 100  &  100 & 100  & 21 & 23 & 3 & 3 \\
&  \multicolumn{7}{c}{\bf Scenario 2} \\
LASSO & 94 &94  & 25 & 23 &  90&  90 & 24\\
OALASSO  & 98  &  98 & 86  & 83 & 64 & 65 & 20 \\
Yang's method & 100  &  100 & 100  & 1.5 & 2 & 1 & 3 \\
&  \multicolumn{7}{c}{\bf Scenario 3} \\
LASSO & 84 &87  & 23 & 23 &  95&  94 & 24\\
OALASSO  & 96  &  96 & 88  & 89 & 84 & 83 & 20 \\
Yang's method & 100  &  100 & 100  & 11 & 2 & 3 & 3 \\
\hline
\end{tabular}
\end{table}

\begin{table}[H]
\small\sf\centering
\caption{Distributions of common covariates from the two samples. }  
\begin{tabular}{l c c } 
\hline
\textbf{} & \textbf{TIO}& \textbf{LFS} \\ [0.1ex]  
\textbf{Covariates} & \textbf{N (mean)}  & \textbf{ N (mean)} \\
\hline 
Sample size & 35916 &  89102 \\
Born in Canada &   30867 (85.94$\%$)
  & 72048(71$\%$)
  \\
  Landed immigrant or permanent resident &   149(41$\%$)	& 15277(26.36$\%$)
  \\
  Sex (1:Male) & 10298(28.67$\%$)&	43415(49.35$\%$)
\\
Rural/Urban indicator (1: rural) & 4395(12.24$\%$)&	14119(8.25$\%$)
\\
Education  &  &  \\
~~~At least High school diploma or equivalency &  35588(99.08$\%$) &	74288(85.46$\%$)  \\
~~~At least Trade certificate or diploma &  32192(89.63$\%$)	& 51036(59.60$\%$) \\
~~~At least College - cegep -other non university certificate  &  30568(85.10$\%$)	& 41038(50.06$\%$)  \\
~~~At least University certificate or diploma below bachelor & 23544(65.55$\%$) &	22826(30.50$\%$) \\
~~~At least Bachelor degree &  21299(59.30$\%$) &	13865(15.56$\%$)  \\
~~~At least University degree above bachelor &  10118(28.17$\%$) & 6526(8.97$\%$)  \\

Indigenous identity flag &  1047(2.92$\%$) &	3689(2.48$\%$)
 \\
 \hline 
\end{tabular}
\label{table:char1} 
\end{table}

 \begin{table}[H]
\small\sf\centering
\caption{Distributions of common covariates from the two samples. }  
\begin{tabular}{l c c } 
\hline
\textbf{} & \textbf{TIO}& \textbf{LFS} \\ [0.1ex]  
\textbf{Covariates} & \textbf{N (mean)}  & \textbf{ N (mean)} \\
\hline
Province  &  &  \\
~~~ Newfoundland and Labrador & 328(0.91$\%$)&	2965(1.41$\%$)\\
~~~ Prince Edward Island & 161(0.45$\%$) &	325(0.42$\%$) \\
~~~ Nova Scotia  & 1762(4.91$\%$)&	4695(2.61$\%$) \\
~~~ New Brunswick  & 794(2.21) & 4727(2.04$\%$)  \\
~~~ Quebec  & 5861(16.32$\%$) &16455(22.85$\%$) \\
~~~ Ontario & 17177(47.83$\%$) &	24978(39.53$\%$) \\
~~~ Manitoba & 922(2.57$\%$) &	7607(3.33$\%$) \\
~~~ Saskatchewan & 890(2.48$\%$) &	6104(2.87$\%$) \\
~~~ Alberta & 2875(8$\%$)& 9265(11.48$\%$) \\
~~~ British Columbia  & 5146(14.33$\%$) &	9981(13.37$\%$) \\
Age group in increments of 10 & & \\
~~~  15-24 &  1113(3.10$\%$) & 	10902(14.15$\%$)    \\
~~~  25-34 &   6162(17.16$\%$) &	12336(16.83$\%$)  \\
~~~  35-44 &   8554(23.82$\%$)&	13573(16.16$\%$)    \\
~~~  45-54 &    7309(20.35$\%$)&	13912(15.11$\%$)    \\
~~~  55-64 &   7111(19.80$\%$)& 6496(16.62$\%$)   \\
~~~ 65+ & 5667(15.78$\%$) &	1883(21.10$\%$) \\
\hline 
\end{tabular}
\label{table:char2} 
\end{table}

\end{appendix}

\end{document}